\documentclass[twocolumn,superscriptaddress,pra,aps]{revtex4-2}

\usepackage{amsfonts}
\usepackage{amssymb}
\usepackage{amsmath}
\usepackage{graphicx}
\usepackage{epsfig}
\usepackage{color,xcolor}
\usepackage{orcidlink}
\setcitestyle{numbers,square,citesep={,\kern-.30em}}
\hypersetup{
	colorlinks=true,    
	pdfborder={0 0 0},
	citecolor=blue,
	urlcolor=blue,    
	linkcolor=blue,
}

\begin{document}
	
	\title{Flat bands and bound states in the continuum in non-Hermitian pseudospin lattices}
	
	\author{X. Y. Chen}
	\author{C. H. Zhang}
	\email{zhangchuhang@mail.nankai.edu.cn}

	\affiliation{School of Physics, Nankai University, Tianjin 300071, China}
	
	\begin{abstract}
		
	Flat bands support compact localized states (CLSs), whereas bound states in the continuum (BICs) remain spatially localized despite having energies embedded in the continuum of extended states. Here, we investigate the connection between these phenomena in a class of one-dimensional non-Hermitian lattices whose Bloch Hamiltonians take the pseudospin form \(H(k)=\mathbf d(k)\cdot\mathbf S\). For integer pseudospin, the Hamiltonian intrinsically supports a zero-energy flat band and its associated compact localized states. When the effective pseudospin field has real transverse components and a purely imaginary longitudinal component, increasing non-Hermiticity compresses the dispersive bands toward the intrinsic zero-energy flat band, enabling their spectral overlap within a finite parameter range. The CLSs thus become embedded in the dispersive continuum and emerge as BICs. Adding a momentum-dependent Hermitian component along the gain–loss direction generally removes this spectral overlap without destroying the CLSs. We further extend this spectral mechanism to integer-pseudospin-\(n\) systems. These findings establish a connection between intrinsic flat-band localization and BIC formation through non-Hermitian spectral reconstruction.
		
	\end{abstract}
	
	\maketitle
	
	\section{Introduction}
	
	Flat bands constitute a unique class of energy bands characterized by the absence of dispersion, leading to a vanishing group velocity throughout the Brillouin zone \cite {LeykamAdvPhysX18,LeykamAPLPhotonics18}. The resulting macroscopic degeneracy provides a natural platform for a variety of remarkable phenomena, including compact localization \cite% 
	{FlachEurophysLett14,XiaoPhysRevLett17,GneitingPhysRevB18}, enhanced interaction effects \cite%
	 {CaoNature18,YankowitzScience19,BalentsNatPhys20}, unconventional transport \cite% 
	 {LeykamPhysRevA12,LeykamPhysRevB13,CaceresAravenaPhysRevA19,CaceresAravenaPhysRevA20}, and unusual fractional topological phases \cite%
	  {BermudezPhysRevLett09,SunPhysRevLett11,TangPhysRevLett11,NeupertPhysRevLett11}. A unique feature of flat-band systems is the existence of compact localized states (CLSs). Their wave functions remain confined within a finite number of unit cells, providing a direct manifestation of the intrinsic localization of flat-band systems. This compact localization enables robust wave confinement and manipulation without diffraction. Flat-band physics has consequently been explored. 
	  Flat bands usually originate from carefully engineered lattice geometries and interference effects, as exemplified by Lieb \cite%
	   {GuzmanSilvaNewJPhys14,ZhangPhysRevA19}, kagome \cite {MasumotoNewJPhys12,ChernPhysRevA14}, dice \cite {VidalPhysRevLett98,KolovskyPhysRevB18} and pyrochlore \cite {ChalkerPhysRevB10,ShuklaPhysRevB18}. They have been realized across a diverse range of physical platforms, spanning condensed-matter systems \cite {LiuChinPhysB14}, photonic structures \cite%
	   {MukherjeePhysRevLett15,VicencioPhysRevLett15,RechtsmanNatPhoton13,BiondiPhysRevLett15,VicencioPobleteAdvPhysX21}, atomic systems \cite {TaieSciAdv15,LeungPhysRevLett20}, and artificial lattice systems \cite {ChaseMayoralPhysRevB23}, underscoring their broad applicability across different physical settings.
	   	
	Another prominent class of localized states is the bound states in the continuum (BICs), which remain spatially confined despite being embedded within the continuum spectrum of extended states. Since their prediction by von Neumann and Wigner \cite {vonNeumannPhysZ29}, BICs have evolved from a fundamental concept in quantum mechanics into a universal wave phenomenon observed in photonics \cite%
	 {DoelemanNatPhotonics18,YuOptica19,KoshelevScience20,KangNatRevPhys23}, acoustics \cite%
	  {FengPhysRevAppl23,HuangAdvSci22,HuangNatCommun21}, electronics \cite {CapassoNature92}, mechanics \cite {ChenNewJPhys16}, and magnonics \cite {YangPhysRevLett20}. 
	A variety of mechanisms have been proposed to realize BICs, including parameter tuning \cite {RamezaniPhysRevA17,MonticonePhysRevLett14}, chiral symmetry \cite {MurPetitPhysRevB14,MukherjeePhysRevLett15}, inverse design \cite {MolinaPhysRevLett12,CorrielliPhysRevLett13}, and parity-time ($\mathcal{PT}$) symmetry \cite {FengLightSciAppl26}. These mechanisms have been experimentally observed and demonstrated across a broad range of optical platforms, including photonic crystal slabs \cite{YinNature20,HwangNatCommun21}, waveguide arrays \cite{WeimannPhysRevLett13}, metasurfaces \cite{KoshelevPhysRevLett18}, and other photonic systems \cite{LonghiPhysRevA08,LonghiPhysRevA09,PlotnikPhysRevLett11}, highlighting the versatility of these platforms for realizing and probing BICs. Flat-band CLSs qualify as BICs only when their eigenvalues lie within the extended-state spectrum in the full
	complex-energy plane. This requirement motivates examining how the dispersive spectrum evolves while compact localization is preserved.
	
	Previous studies have established connections between
	non-Hermitian flat bands, compact localization, and exceptional
	degeneracies. Ramezani demonstrated a real-energy flat band
	embedded in dispersive bands at an exceptional point of a
	$\mathcal{PT}$-symmetric photonic lattice
	\cite{RamezaniPhysRevA17}.
	Leykam, Flach, and Chong investigated non-Hermitian flat bands
	supported by bipartite sublattice symmetry, including compact
	localized states and exceptional band crossings
	\cite{Leykam2017}.
	Jin showed that balancing synthetic magnetic flux and
	non-Hermiticity yields a real-energy flat band whose compact
	states form BICs where it intersects the dispersive bands
	\cite{Jin2019}.
	Building on these results, we ask under what conditions
	varying non-Hermiticity brings the dispersive spectrum into
	contact with an existing zero-energy flat band while
	preserving compact localization.
	In particular, we seek to determine when this contact
	persists over a finite interval of non-Hermiticity and
	whether the same conditions extend to higher integer
	pseudospins.

    In this work, we establish a systematic non-Hermitian pseudospin framework based on the generic Bloch Hamiltonian $H(k) = \boldsymbol{d}(k)\cdot\boldsymbol{S}$, in which integer pseudospin representations intrinsically support a zero-energy flat band. We show that non-Hermiticity controls the spectral separation between the intrinsic zero-energy flat band and the dispersive branches, producing a finite interval of spectral embedding. We express the zero-energy contact condition through the orthogonality and equal magnitudes of the real and imaginary effective-field vectors. It demonstrates how an additional Hermitian coupling restricts the embedding to an isolated critical parameter value while preserving compact localization. We further extend the spectral analysis to arbitrary integer-pseudospin representations.
		
	\section{Model}
	
\begin{figure}[t]
	\centering
	\includegraphics[bb=0 10 700 400,width=8.8cm]{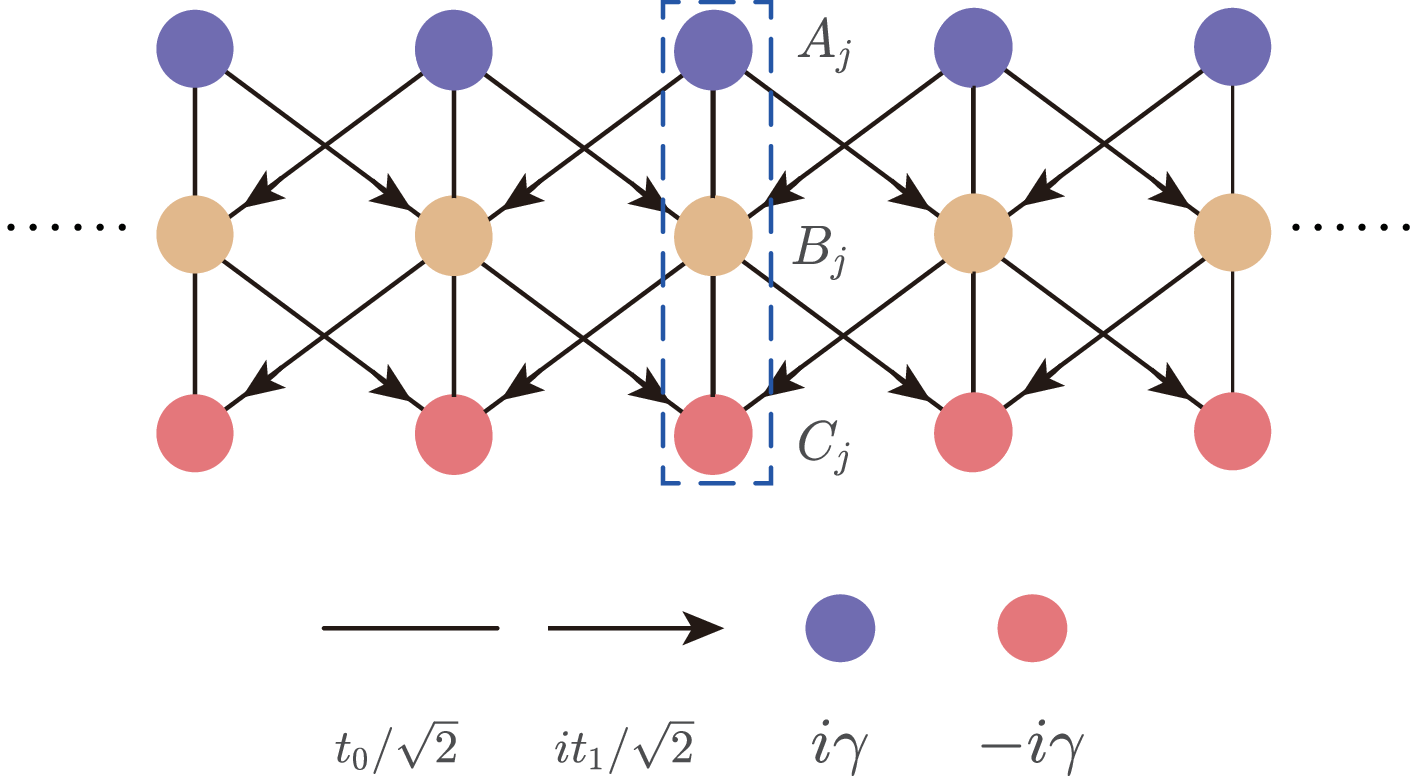}
	
	\caption{Schematic of the one-dimensional non-Hermitian pseudospin-1 lattice. The lattice consists of three sublattices, $A$, $B$, and $C$, carrying gain, passive, and loss potentials, respectively. Black arrows denote nonreciprocal couplings, while black solid lines indicate reciprocal couplings.}
	
	\label{fig1}
\end{figure}

	Figure~\ref{fig1} schematically depicts the quasi-one-dimensional periodic non-Hermitian lattice considered in this work. The lattice consists of unit cells (blue dashed boxes), each containing three sublattice sites labeled $A$, $B$, and $C$, which are denoted by $A_j$, $B_j$, and $C_j$ in the $j$-th unit cell. The $A$, $B$, and $C$ sublattices correspond to gain, passive, and loss sites with balanced gain and loss rate $\gamma$. The system is described by the Hamiltonian:
	
	\begin{align}
		H=&\sum_n \bigg[\frac{t_0}{\sqrt{2}}
		\left(a_n^\dagger b_n+b_n^\dagger c_n\right)
		+\frac{i t_1}{\sqrt{2}}\sum_{\delta=\pm1}
		\big(
		b_n^\dagger a_{n+\delta}\notag
		\\
		&
		+c_n^\dagger b_{n+\delta}\big)+\mathrm{H.c.}
		+i\gamma
		\left(a_n^\dagger a_n-c_n^\dagger c_n\right)\bigg],\label{eq:H_real}
	\end{align}

	where
	$a_n^\dagger$ ($a_n$),
	$b_n^\dagger$ ($b_n$),
	and
	$c_n^\dagger$ ($c_n$)
	are the creation (annihilation) operators that satisfy the periodic boundary condition $a_{N+1}=a_{1}$, $b_{N+1}=b_{1}$, and $c_{N+1}=c_{1}$, respectively. The system parameters ($\gamma$, $t_0$, $t_1$) are all considered to be positive real values without loss of generality.
	
Applying the Fourier transformation $(a_n,b_n,c_n)^T=\sum_k
e^{ikn}(a_k,b_k,c_k)^T/{\sqrt{N}}$, where $k=2\pi m/N$ $(m\in[1,N])$ denotes the crystal momentum, the Hamiltonian is rewritten in momentum space as $H=\sum_k\Psi_k^\dagger H(k)\Psi_k$,
where the Bloch Hamiltonian $H(k)$ reads
\begin{equation}
	H(k)=
	\begin{pmatrix}
		i\gamma & v(k) & 0\\
		v^*(k) & 0 & v(k)\\
		0 & v^*(k) & -i\gamma
	\end{pmatrix},
	\label{eq:Hform}
\end{equation}
with the complex hopping amplitude of the Hamiltonian $v(k)=\left(t_0-2it_1\cos k\right)/\sqrt2$.
	
	Having established the Bloch Hamiltonian, we now reformulate the model within a pseudospin framework. To this end, we introduce the pseudospin-1 operators $\boldsymbol S=(S_x,S_y,S_z)$, whose matrix representations are
\begin{equation}
	\begin{aligned}
		S_x &=
		\frac{1}{\sqrt{2}}
		\begin{pmatrix}
			0 & 1 & 0\\
			1 & 0 & 1\\
			0 & 1 & 0
		\end{pmatrix},
		\qquad
		S_y =
		\frac{1}{\sqrt{2}}
		\begin{pmatrix}
			0 & -i & 0\\
			i & 0 & -i\\
			0 & i & 0
		\end{pmatrix},
		\\[4pt]
		S_z &=
		\begin{pmatrix}
			1 & 0 & 0\\
			0 & 0 & 0\\
			0 & 0 & -1
		\end{pmatrix}.
	\end{aligned}
	\label{eq:spin1_matrices}
\end{equation}
	By expanding the Bloch Hamiltonian over this operator basis, $H(k)$ can be expressed in the compact form
	\begin{equation}
		H(k)=\boldsymbol d(k)\cdot\boldsymbol S,
	\end{equation}
	where the effective pseudospin field is given by $\boldsymbol d(k)=\left(t_0,\,2t_1\cos k,\,i\gamma\right)$. In this representation, the hopping amplitudes are determined by the pseudospin components $d_x=t_0$ and $d_y=2t_1\cos k$, while the non-Hermitian gain–loss term contributes to the component $d_z=i\gamma$. This representation provides a unified framework for characterizing the interplay between lattice couplings and non-Hermiticity, and serves as the basis for the subsequent analysis of the spectral properties and bound states in the continuum \cite {BerciouxPhysRevA09,DoraPhysRevB11}. 

    The eigenvalues of this pseudospin-$1$ Hamiltonian are
    \begin{equation}
    	E(k)=0,
    	\pm
    	\sqrt{
    		d_x^2+d_y^2+d_z^2
    	},
    \end{equation}
    yielding 
    \begin{equation}
    	E(k)
    	=0,
    	\pm
    	\sqrt{
    		t_0^2
    		+
    		4t_1^2\cos^2k
    		-
    		\gamma^2
    	}. 
    	\end{equation}
    	The zero-energy flat band remains pinned at $E=0$, reflecting the intrinsic spectral feature of the pseudospin-1 structure. In contrast, the dispersive branches will undergo band compression induced by the imaginary pseudospin component $d_z = i\gamma$. With increasing non-Hermiticity, the dispersive spectrum is gradually driven from the real energy into the complex energy.

       The dispersive bands are compressed toward the intrinsic zero-energy flat band, which remains protected by the $\mathcal{PT}$ symmetry. Specifically, the Hamiltonian satisfies $(\mathcal{PT})H(k)(\mathcal{PT})^{-1}$=$H(k)$,
     where the parity operator is defined as
     \begin{equation}
     	\mathcal{P}=
     	\begin{pmatrix}
     		0&0&1\\
     		0&1&0\\
     		1&0&0
     	\end{pmatrix},
     \end{equation}
     and the time-reversal operator satisfies $\mathcal{T}i\mathcal{T}^{-1}=-i$. The zero-energy flat band is enforced by the integer-pseudospi$n$ form of the Hamiltonian. In addition, the present model possesses $\mathcal{PT}$ symmetry, which constrains the dispersive spectrum and its transition between real and complex eigenvalues. We decompose the complex effective field into two real vectors,
     \begin{equation}
     	\boldsymbol d(k)=\boldsymbol a(k)+i\boldsymbol b(k),
     	\qquad \boldsymbol a(k),\boldsymbol\in\mathbb R^3,
     \end{equation}
     which determine the Hermitian and anti-Hermitian parts of the Bloch Hamiltonian, respectively. For the present model 
     $\boldsymbol a(k)=(t_0,2t_1\cos k,0), \boldsymbol b(k)=(0,0,\gamma).$ These vectors are orthogonal throughout the Brillouin zone, so that $\boldsymbol d(k)\cdot\boldsymbol d(k)
     =|\boldsymbol a(k)|^2-|\boldsymbol b(k)|^2
     =t_0^2+4t_1^2\cos^2k-\gamma^2.$ Thus, the effective-field orthogonality eliminates the imaginary cross term in the squared dispersive eigenvalues.
     
\begin{figure}[t]
	\centering
	\includegraphics[width=\columnwidth]{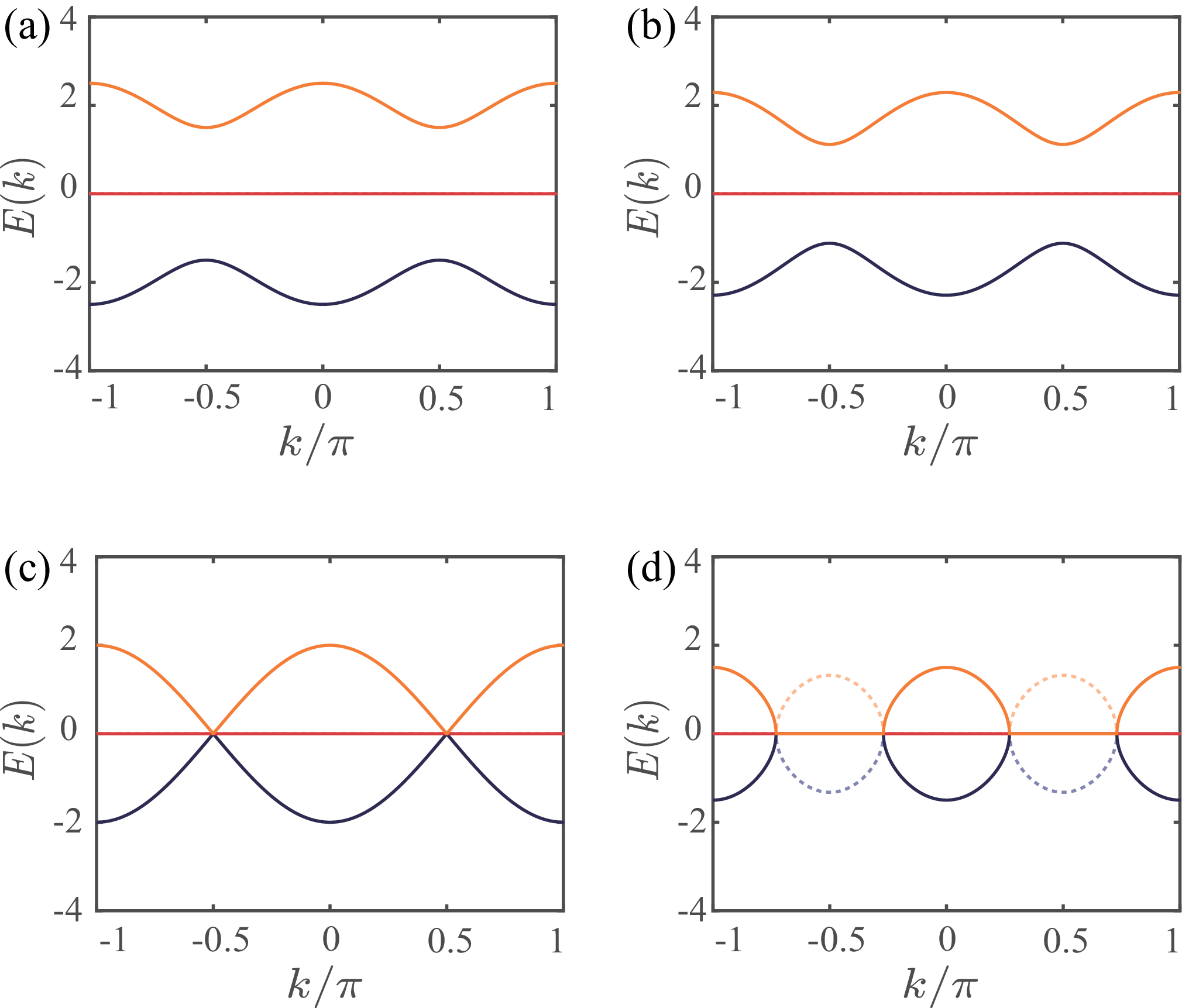}
	\caption{Energy spectra of the lattice for different values of the non-Hermitian parameter $\gamma$. Solid (dashed) curves represent the real (imaginary) parts of the energy spectrum. (a)–(d) Band structures of the non-Hermitian lattice shown in Fig.~\ref{fig1} for $\gamma=0$, $1$, $1.5$, and $2$, respectively. The coupling amplitudes are fixed at $t_0=1.5$ and $t_1=1$. The Brillouin zone is $k\in[-\pi,\pi]$.} \label{fig2}
\end{figure}

     The evolution of the spectrum with increasing non-Hermiticity under $\mathcal{PT}$ symmetry is illustrated in Fig.~\ref{fig2}. For $\gamma=0 $ [Fig.~\ref{fig2}(a)], the system reduces to the Hermitian limit, where the dispersive bands exhibit a conventional real-valued spectrum and remain separated from the zero-energy flat band. As the gain–loss strength is increased, the system remains in the exact $\mathcal{PT}$-symmetric phase when $\left|\boldsymbol b(k)\right|^2
     <
     \left|\boldsymbol a(k)\right|^2$ under which all eigenvalues remain real, as illustrated in Fig.~\ref{fig2}(b). In this regime, the dispersive branches gradually shrink toward the flat band owing to the increasing imaginary component $d_z$ of the pseudospin field. At a critical gain–loss strength $\gamma=1.5$, the dispersive bands touch the zero-energy flat band, resulting in a gap closing [Fig.~\ref{fig2}(c)]. At the critical point,
     $\left|\boldsymbol b(k)\right|
     =
     \left|\boldsymbol a(k)\right|$, and the three energy bands coalesce at zero energy, giving rise to third-order exceptional points (EP3s). Upon further increasing $\gamma$, such that $\left|\boldsymbol b(k)\right|^2
     >
     \left|\boldsymbol a(k) \right|^2$, the system enters the broken $\mathcal{PT}$-symmetric phase, where complex spectral regions emerge in the Brillouin zone. This spectral evolution is consistent with the general behavior of $\mathcal{PT}$-symmetric systems, in which the exact and broken phases are separated by exceptional points (EPs)
     \cite{BenderPhysRevLett98,DembowskiPhysRevLett01}.

	\section{Compact Localized States and bound states in the continuum}
	
Compact localization and spectral embedding are distinct requirements for a bound state in the continuum \cite{HsuNatRevMater16}. We first examine the eigenstates of the non-Hermitian pseudospin-1 lattice under open boundary conditions (OBCs) and explicitly construct the compact localized states (CLSs) associated with the intrinsic flat band.
	
	We now transform the Bloch Hamiltonian into its corresponding real-space representation. The wave function is written as
\begin{equation}
	|\psi\rangle = \sum_j \left(
	\psi_{A_j}|A_j\rangle
	+ \psi_{B_j}|B_j\rangle
	+ \psi_{C_j}|C_j\rangle
	\right),
\end{equation}
	which obeys the Schrödinger equation
	$i\frac{d}{dt}|\psi\rangle=H|\psi\rangle$.
	By projecting this equation onto the three sublattices within each unit cell, we obtain the real-space Schrödinger equations for the non-Hermitian pseudospin-$1$ lattice shown in Fig.~\ref{fig1}:
	
	\begin{eqnarray}
		i\dot{\psi}_{A_j}
		&=& i\gamma\psi_{A_j}
		+\frac{t_0}{\sqrt{2}}\psi_{B_j}
		-i\frac{t_1}{\sqrt{2}}
		(\psi_{B_{j-1}}+\psi_{B_{j+1}}),
		\label{A}
		\\
		i\dot{\psi}_{B_j}
		&=&
		\frac{t_0}{\sqrt{2}}
		(\psi_{A_j}+\psi_{C_j})
		+i\frac{t_1}{\sqrt{2}}
		(\psi_{A_{j-1}}+\psi_{A_{j+1}}\notag \\
		&&
		-\psi_{C_{j-1}}-\psi_{C_{j+1}}),
		\label{B}
		\\
		i\dot{\psi}_{C_j} &=& \frac{t_0}{\sqrt{2}}\psi_{B_j} +i\frac{t_1}{\sqrt{2}} (\psi_{B_{j-1}}+\psi_{B_{j+1}})-i\gamma\psi_{C_j}. \label{C} 
	\end{eqnarray}
	
	From Eqs.~(\ref{A}--\ref{C}), it is obvious that the $B$ sublattice acts as an intermediate channel that mediates the coupling between neighboring unit cells, whereas the $A$ and $C$ sublattices carry the balanced gain and loss. Solving the zero-energy eigenvalue problem $(E=0)$ analytically yields compact localized eigenstates associated with the intrinsic flat band.
	
The intrinsic zero-energy flat band supports compact
localized states at both the boundary and in the bulk.
The corresponding boundary and bulk CLSs are
\begin{equation}
	\psi_{\mathrm{CLS}}^{\mathrm{edge}}
	=
	\left(
	\boldsymbol{\phi}_2,
	\boldsymbol{\phi}_1,
	0,0,0,\ldots
	\right)^{T},
	\label{CLS_edge}
\end{equation}
and
\begin{equation}
	\psi_{\mathrm{CLS}}^{\mathrm{bulk}}
	=
	\left(
		0,0,0,
	\boldsymbol{\phi}_1,
	\boldsymbol{\phi}_2,
	\boldsymbol{\phi}_1,
		0,0,0,\ldots
	\right)^{T},
	\label{CLS_bulk}
\end{equation}
respectively. Here, the three-component row vectors
are defined as
$\boldsymbol{\phi}_1=(1,0,1)$ and
$\boldsymbol{\phi}_2=
(it_0/t_1,\sqrt{2}\gamma/t_1,-it_0/t_1)$.

\begin{figure}[t]
	\centering
	\includegraphics[trim=108bp 0bp 0bp 6bp,clip,width=\columnwidth]{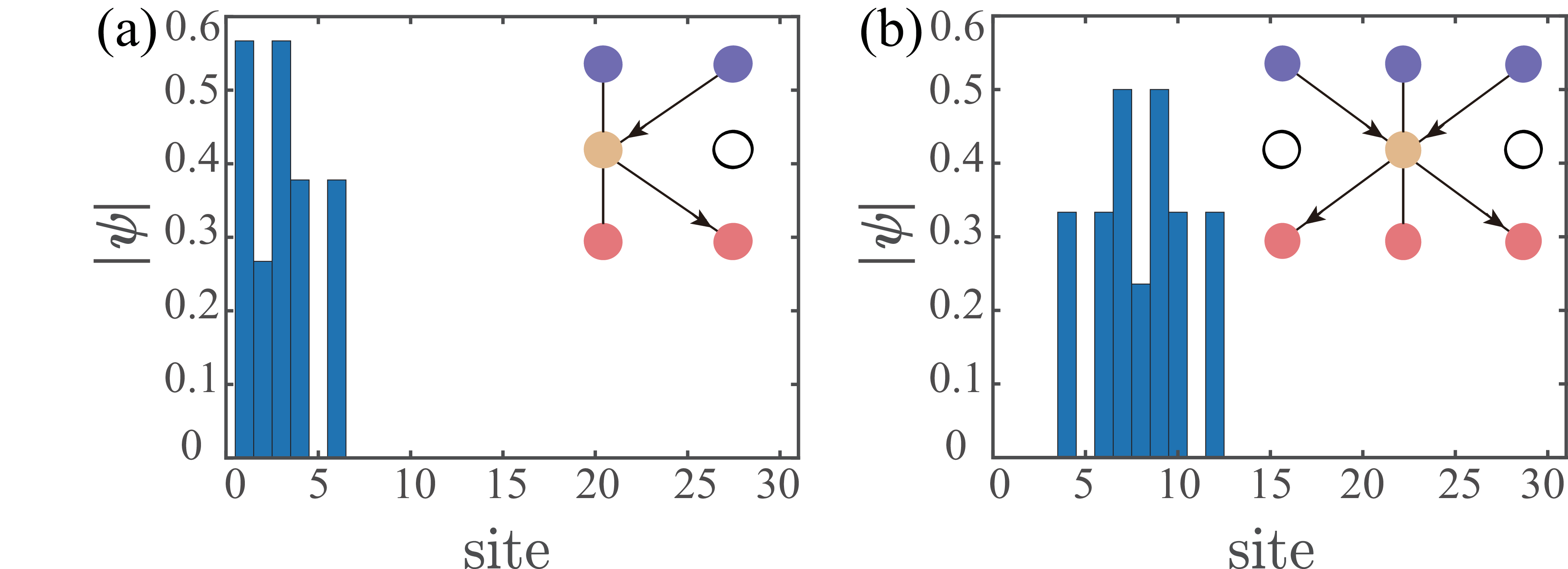}
	\caption{(a) Edge CLS in a one-dimensional non-Hermitian pseudospin-1 lattice under OBCs, with the boundary unit cell decoupled from the bulk. (b) Bulk CLS under OBCs, with the localized bulk region decoupled from the boundary and the rest of the lattice. The system consists of $N = 10$ unit cells. The hopping parameters are $t_0 = 1.5$, $t_1 = 1$, with non-Hermiticity parameter $\gamma=0.5$.
	} \label{fig3}
\end{figure}

The states in Eqs.~\ref{CLS_edge} and \ref{CLS_bulk} occupy two and three unit cells,
respectively. Figure~\ref{fig3} displays their normalized amplitude profiles
for $N=10$ and $\gamma=0.5$. These examples establish compact
localization, but not spectral embedding: at the parameters of
Fig~\ref{fig3}, the dispersive bands remain separated from $E=0$.
The same analytical states remain exact zero-energy eigenstates
as $\gamma$ varies, although their relative amplitudes change.

 To determine when these states become spectrally embedded, we compare their zero-energy eigenvalue with the dispersive Bloch spectrum of the infinite periodic lattice. When the flat-band energy coincides with a dispersive Bloch eigenvalue, the CLSs become spectrally embedded and can be identified as BICs. Because the dispersive eigenvalues can be complex, this comparison must involve both their real and imaginary parts. We therefore define the minimum separation between the flat-band energy and the two dispersive branches in the complex-energy plane as
\begin{equation}
	\begin{aligned}
		\Delta_{\min}(k,\gamma)
		&= \left|E_{\nu}(k,\gamma)-E_0\right| \\
		&= \sqrt{\left|
			t_0^2+4t_1^2\cos^2 k-\gamma^2
			\right|}.
	\end{aligned}
	\label{eq:complex_energy_separation}
\end{equation}

 Here, \(E_0=0\) denotes the flat-band energy, while \(E_{\pm}(k,\gamma)\) denote the eigenvalues of the two dispersive branches at Bloch momentum \(k\) and gain--loss strength \(\gamma\).
 
\begin{figure}[t]
	\centering
	\includegraphics[trim=0bp 0bp 0bp 212bp,clip,width=\columnwidth]{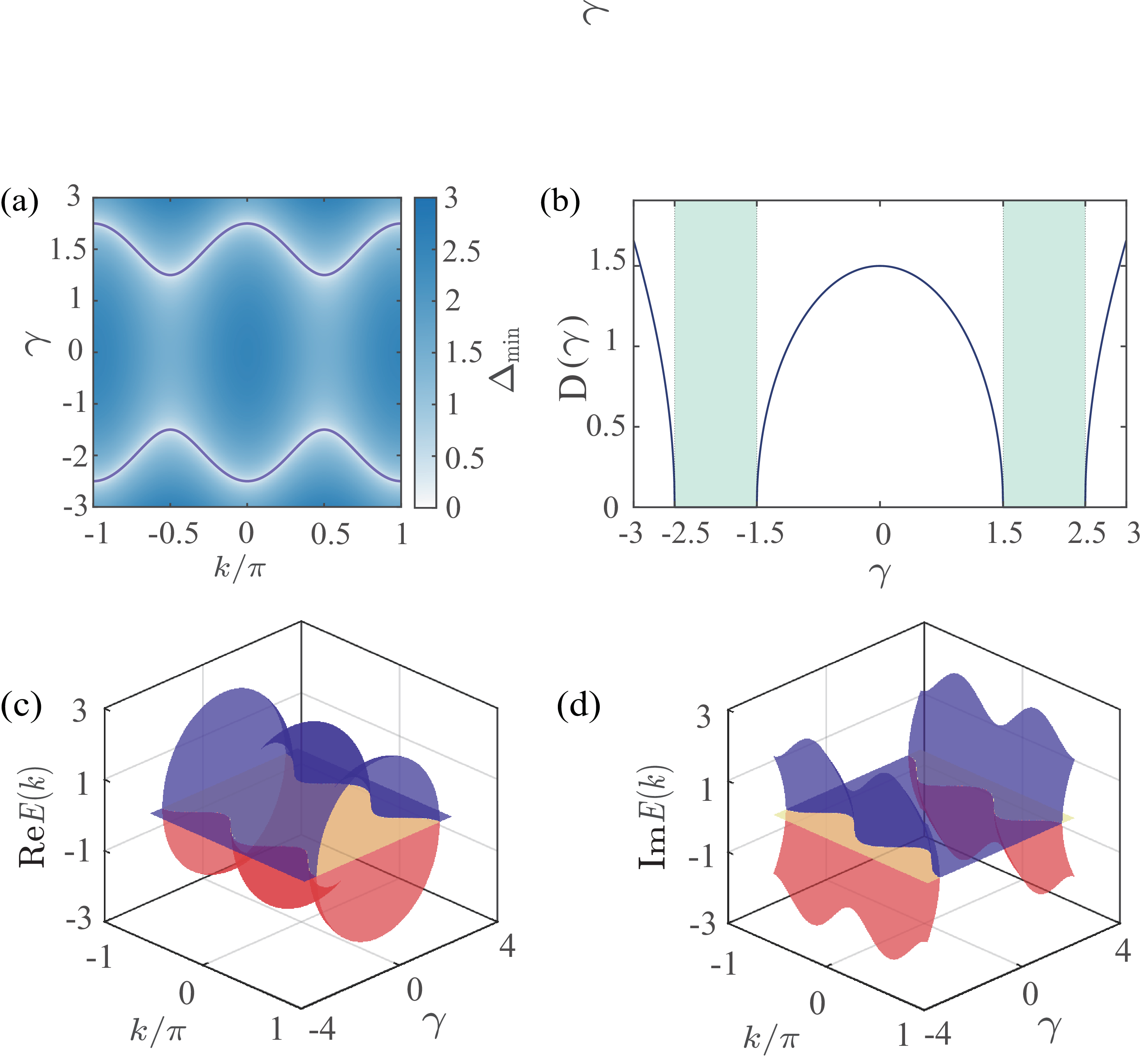}
	\caption{Evolution of the energy spectrum with the non-Hermitian parameter $\gamma$.  (a) Color map of the complex-energy separation
		$\Delta_{\min}(k,\gamma)$ as a function of Bloch momentum $k$
		and gain--loss strength $\gamma$.
		 (b) Global separation
		$D(\gamma)=\min_{k\in\mathrm{BZ}}\Delta_{\min}(k,\gamma)$.
		The shaded regions indicate $D=0$, corresponding to
		$1.5\leq|\gamma|\leq2.5$, with the vertical lines marking
		the endpoints of these intervals. (c), (d) Real and imaginary parts of the band structure of the non-Hermitian lattice shown in Fig.~\ref{fig1} as functions of crystal momentum $k$ and $\gamma$. The parameters are $t_0=1.5$ and $t_1=1$, with $\gamma\in[-3,3]$ and $k\in[-\pi,\pi]$.}  \label{fig4}
\end{figure}

 The exact CLSs remain pinned at zero energy as $\gamma$ varies,
 whereas the dispersive branches shift with the gain--loss
 strength. When
 $\gamma^2=t_0^2+4t_1^2\cos^2 k$,
 the separation $\Delta_{\min}(k,\gamma)$ vanishes, and the dispersive
 eigenvalues reach the flat-band energy without destroying
 the compact localization of these states.
 The purple curves in Fig.~\ref{fig4}(a) trace this contact condition:
 for each $\gamma$ admitting a real solution for $k$,
 the zero-energy CLSs are spectrally embedded in the
 dispersive spectrum.
 
 The range of $\gamma$ supporting this embedding follows
 from the requirement $0\leq\cos^2 k\leq1$, which gives
 \begin{equation}
 	\begin{aligned}
 		t_0\leq|\gamma|\leq\sqrt{t_0^2+4t_1^2}.
 	\end{aligned}
 	\label{eq:rangegamma}
 \end{equation}
 
 This range is captured by the global separation
 $D(\gamma)=\min_{k\in\mathrm{BZ}}\Delta_{\min}(k,\gamma)$,
 shown in Fig.~\ref{fig4}(a).
 For $t_0=1.5$ and $t_1=1$, $D$ vanishes throughout
 $1.5\leq|\gamma|\leq2.5$, corresponding to the shaded
 intervals. Within their interiors, the compact states
 are embedded in the dispersive spectrum, while the
 endpoints mark the onset and termination of spectral
 contact. 
 \label{key}
	As a consequence, the intrinsic flat band is no longer spectrally isolated but becomes embedded within the continuum formed by the two gapless dispersive bands as shown in Fig.~\ref{fig4}(c) and Fig.~\ref{fig4}(d). Since the associated eigenstates retain their compact localization, the embedded flat-band states satisfy the defining characteristics of BICs \cite {XiaoPhysRevLett17,HsuNatRevMater16}, namely, spatially localized eigenstates with eigenenergies embedded in the continuum spectrum. Such states exhibit intrinsically suppressed radiation leakage and exceptionally long lifetimes, enabling a wide range of applications, from high-$Q$ photonic resonators and ultrasensitive sensing to beam manipulation and wavelength-selective filtering \cite{KoshelevPhysRevLett18}.

	\section{Effective-field conditions for spectral embedding}
	
The preceding results establish a finite interval of
non-Hermiticity over which the zero-energy CLSs are
embedded in continuum spectrum.We now examine how this interval
depends on the effective-field structure while compact
localization is retained. To this end, we keep
$d_x=t_0$ and $d_y=2t_1\cos k$ unchanged and introduce
a Hermitian contribution to the longitudinal component,
\begin{equation}
	d_z=2t_2\cos k+i\gamma.
	\label{19}
\end{equation}
where $2t_2\cos k$ is a momentum-dependent Hermitian contribution and $i\gamma$ represents the non-Hermitian gain–loss term. 
This modification preserves the pseudospin-1 form of the
Hamiltonian but changes the conditions under which the
dispersive bands can reach the flat-band energy.This allows us to directly assess how an additional Hermitian contribution to the component $d_z$ affects the band evolution and the formation of BICs.

\begin{figure}[t]
	\centering
	\includegraphics[width=8.8cm]{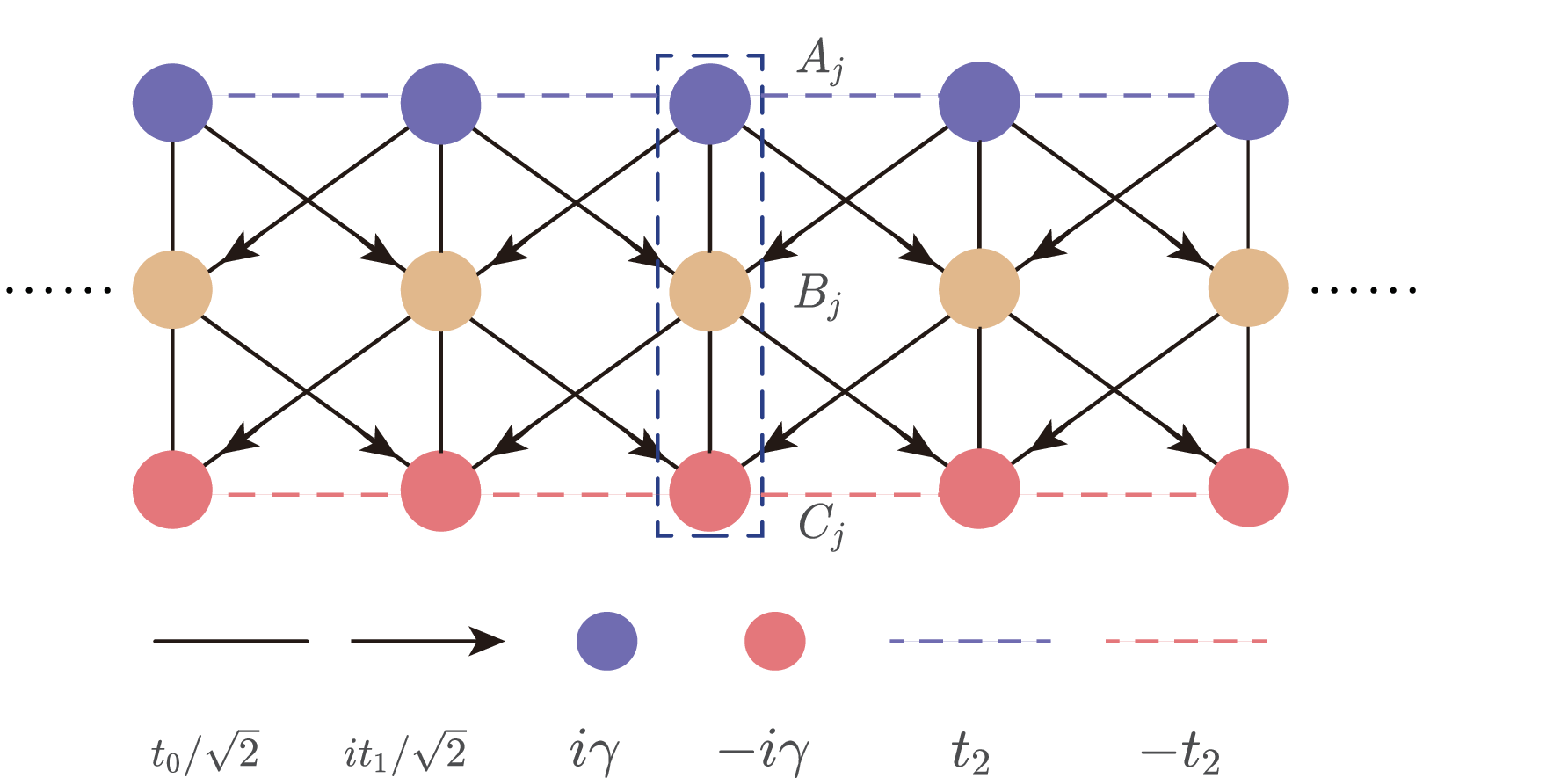}
	\caption{Schematic of the generalized non-Hermitian pseudospin-1 lattice corresponding to $d_z=2t_2\cos k+i\gamma$. The colored dashed lines represent the additional nearest-neighbor intercell couplings introduced by the Hermitian contribution $2t_2\cos k$.}
	\label{fig5}
\end{figure}

	The additional Hermitian contribution $2t_2\cos k$ corresponds to nearest-neighbor couplings between the $A$ and $C$ sublattices of adjacent unit cells, as indicated by the dashed bonds in Fig.~\ref{fig5}. Accordingly, the Bloch Hamiltonian becomes
	
	\begin{equation}
		H(k)=
		\begin{pmatrix}
			2t_2\cos k+i\gamma & v(k) & 0\\
			v^*(k) & 0 & v(k)\\
			0 & v^*(k) & -2t_2\cos k-i\gamma
		\end{pmatrix},
	\end{equation}
where	$v(k)=\left(t_0-2it_1\cos k\right)/\sqrt2$.

	 Although the generalized Hamiltonian still possesses the pseudospin-1 form and therefore preserves the intrinsic zero-energy flat band, $d_z$ of the generalized effective field now contains both Hermitian and non-Hermitian contributions
	\begin{equation}
		d_z=\mathrm{Re}(d_z)+i\,\mathrm{Im}(d_z),
	\end{equation}
	with $\mathrm{Re}(d_z)=2t_2\cos k, \mathrm{Im}(d_z)=\gamma$.
	The dispersive eigenvalues become
	\begin{equation}
		E(k)=0,\quad
		\pm\sqrt{
			d_x^2+d_y^2+
			\left[\mathrm{Re}(d_z)+i\,\mathrm{Im}(d_z)\right]^2
		}.
	\end{equation}
	
	We find this complex form introduces the mixed term $2i\,\mathrm{Re}(d_z)\mathrm{Im}(d_z)$, which directly mixes the Hermitian and non-Hermitian contributions to the  $d_z$ component. As a result, the dispersive eigenvalues acquire a momentum-dependent complex structure, fundamentally modifying the spectral evolution underlying the BIC formation in the original model. In particular, the condition $E(k)=0$ for the dispersive bands now requires both the real and imaginary parts of the radicand to vanish simultaneously. For  $d_z=2t_2\cos k+i\gamma$, this condition reduces to $4\gamma t_2 \cos k = 0,  t_0^2 + 4(t_1^2 + t_2^2)\cos^2 k - \gamma^2 = 0$.
For the modified field,
the two contact conditions become
\begin{equation}
	\begin{aligned}
		4\gamma t_2\cos k &=0,\\
		t_0^2+4(t_1^2+t_2^2)\cos^2 k-\gamma^2&=0.
	\end{aligned}
\end{equation}
For $t_0>0$, $t_2\neq0$, and $\gamma\neq0$, they yield
$ k=\pm\frac{\pi}{2}, |\gamma|=t_0.$
At these points, all three eigenvalues coalesce at zero,
forming third-order exceptional points.
Thus, the finite embedding interval of the original
model collapses to the critical parameter values
$|\gamma|=t_0$.
	
		\begin{figure}[t]
		\centering
		\includegraphics[bb=0 20 950 850,width=8.8cm]{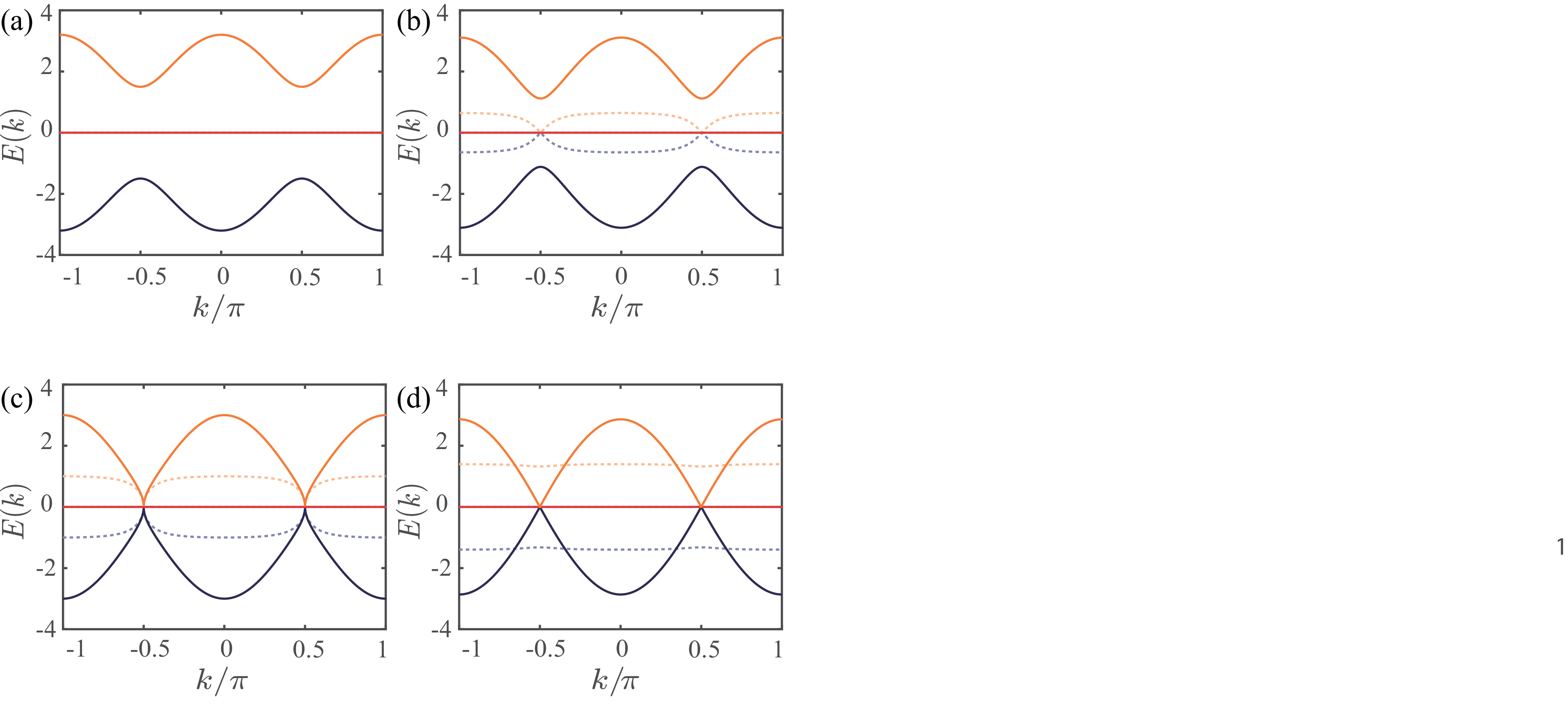}
		\caption{(a)--(d) Corresponding energy spectra upon introducing the Hermitian contribution $2t_2\cos k$ to the $d_z$ component, as illustrated in Fig.~\ref{fig5}, for $\gamma=0$, $1$, $1.5$, and $2$, respectively. The coupling amplitudes are fixed at $t_0=1.5$, $t_1=1$, and $t_2=1$. The Brillouin zone is defined as $k\in[-\pi,\pi]$.}
		\label{fig6}
	\end{figure}
	
	Figure~\ref{fig6} illustrates this restriction of spectral
	contact. In the Hermitian limit $\gamma=0$
	[Fig.~\ref{fig6}(a)], the dispersive bands are real and
	separated from the zero-energy flat band.
	For $0<\gamma<t_0$ [Fig.~\ref{fig6}(b)], the dispersive
	eigenvalues are generally complex, but neither branch
	reaches $E=0$.
	At $\gamma=t_0$ [Fig.~\ref{fig6}(c)], both their real and
	imaginary parts vanish at $k=\pm\pi/2$, producing
	zero-energy contacts with the flat band.
	For $\gamma>t_0$ [Fig.~\ref{fig6}(d)], the dispersive
	eigenvalues at these momenta become
	$E_\pm=\pm i\sqrt{\gamma^2-t_0^2}$.
	Their real parts therefore vanish, whereas their
	nonzero imaginary parts separate them from the flat-band
	energy. Unlike the finite interval in Fig.~\ref{fig4},
	spectral embedding here occurs only at $|\gamma|=t_0$,
	although compact localization persists away from these
	values. Thus, while compact localization is preserved, spectral embedding is restricted to $|\gamma|=t_0$, rather than
	occurring over a finite interval of non-Hermiticity.
	At these critical values, the zero-energy CLSs form BICs;
	away from them, the CLSs remain compact but their energy
	lies outside the dispersive spectrum.
	
	\begin{figure*}[t]
		\includegraphics[width=18cm]{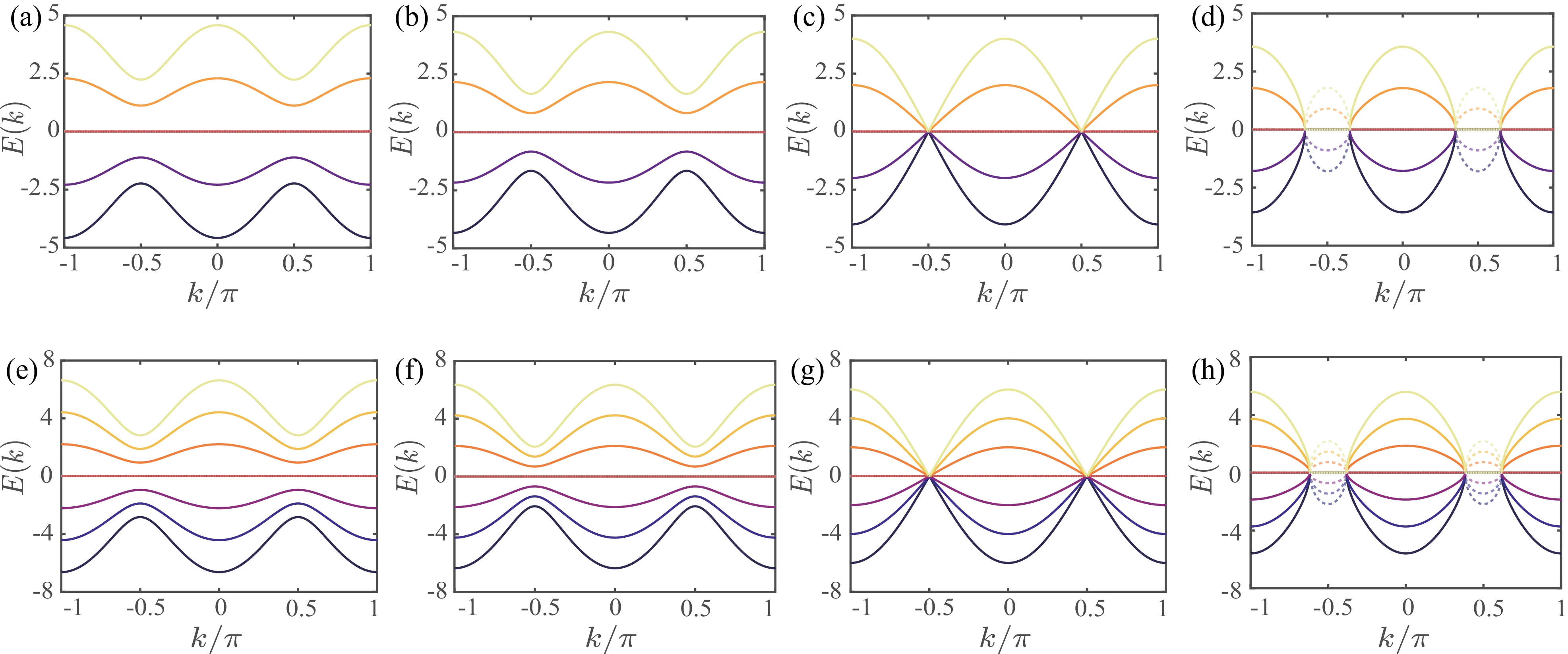}
		\caption{Energy spectra of higher-order pseudospin lattice systems for different values of the non-Hermitian parameter $\gamma$. Solid (dashed) lines indicate the real (imaginary) parts of the energy spectrum. (a)–(d)  Pseudospin-$2$ lattice for $\gamma = 0, 1, 1.5,$ and $2$, respectively. (e)–(h) Pseudospin-$3$ lattice for $\gamma = 0,1, 1.5,$ and $2$, respectively. The coupling parameters are $t_0 = 1.5$ and $t_1 = 1$. The Brillouin zone is defined by $k \in [-\pi, \pi]$.} \label{fig7}
	\end{figure*}
	
	The effective-field structure therefore
	determines whether the preserved CLSs become BICs over
	a finite parameter interval or only at isolated
	parameter values.

	\section{ Generalization to pseudospin-$n$ systems }
	
	The analysis in the previous section establishes that the formation of BICs is not a generic consequence of non-Hermiticity, but instead requires a specific structure of the effective pseudospin field. 

	In particular, the separation between the Hermitian and non-Hermitian contributions to the component $d_z$ constitutes the key condition that enables the dispersive bands to approach the intrinsic zero-energy flat band while leaving the flat-band energy unchanged. This naturally raises a more general question: Is this condition unique to the pseudospin-$1$ lattice, or does it remain valid for arbitrary integer-pseudospin systems?
	
	To answer this question, we extend the present framework to higher-dimensional integer pseudospin representations while preserving the effective pseudospin field  $\boldsymbol{d}(k)=\left(t_0, 2t_1\cos k, i\gamma\right)$.
	For an arbitrary integer pseudospin $n$, the spin-1 generators are promoted to the corresponding $(2n+1)$-dimensional spin operators
	
	\begin{equation}
		\boldsymbol{S}^{(n)} = \left(S_x^{(n)}, S_y^{(n)}, S_z^{(n)}\right),
	\end{equation}
	while the effective pseudospin field is kept unchanged,
	
		\begin{equation}
			\boldsymbol{d}(k) = \left(t_0,\,2t_1\cos k,\,i\gamma\right).
		\end{equation}
		Accordingly, the Bloch Hamiltonian retains the universal pseudospin form
		\begin{equation}
			H^{(n)}(k) = \boldsymbol{d}(k)\cdot \boldsymbol{S}^{(n)}.
		\end{equation}
		Since the effective pseudospin field preserves the same structure identified in Sec.~II, the energy spectrum follows directly from the eigenvalues of the pseudospin-$n$ generators
		\begin{equation}
			E_m(k) = m\sqrt{d_x^2+d_y^2+d_z^2}  \quad(m=-n,\cdots,n),
		\end{equation}
		or equivalently,
		\begin{equation}
			E_m(k) = m \sqrt{t_0^2 + 4t_1^2 \cos^2 k - \gamma^2}.
		\end{equation}
	
	The corresponding band structures for the pseudospin-$2$ and pseudospin-$3$ lattices are presented in Fig.~\ref{fig7}, respectively. Owing to the preserved pseudospin structure $	H^{(n)}(k) = \boldsymbol{d}(k)\cdot \boldsymbol{S}^{(n)}$, the essential spectral features established for the pseudospin-1 lattice remain unchanged. In particular, the intrinsic zero-energy flat band persists for all integer pseudospin representations, while the $\mathcal{PT}$ symmetry of the Hamiltonian ensures the same evolution from the exact to the broken $\mathcal{PT}$-symmetric phase.
	
	As the gain--loss strength increases, the dispersive bands of the pseudospin-$2$ and pseudospin-$3$ lattices are continuously compressed toward the intrinsic zero-energy flat band shown in Figs.~\ref{fig7}(a) and~\ref{fig7}(b), and Figs.~\ref{fig7}(e) and~\ref{fig7}(f), respectively. At the critical non-Hermitian strength [Figs.~\ref{fig7}(c) and~\ref{fig7}(g)], the dispersive bands coalesce with the flat band at zero energy, forming higher-order EPs whose order is determined by the dimension of the pseudospin representation. In particular, the pseudospin-$2$ and pseudospin-$3$ lattices exhibit EP5s and EP7s, respectively, corresponding to the coalescence of all $2n+1$ bands at $E=0$. Upon further increasing the gain–loss strength [Figs.~\ref{fig7}(d) and~\ref{fig7}(h)], the systems enter the broken $\mathcal{PT}$-symmetric phase, where the dispersive bands acquire finite imaginary components while the zero-energy flat band remains pinned at $E=0$. The flat band is thereby embedded in the complex-energy continuum formed by the dispersive bands, while its associated eigenstates retain their compact localization. These results demonstrate that the BIC-forming mechanism identified above extends naturally to higher-dimensional integer-pseudospin representations, with the spectral coalescence occurring at correspondingly higher-order EPs.
	
	The results for pseudospin-$2$ and pseudospin-$3$ indicate that the BIC-forming mechanism identified above is not restricted to the pseudospin-$1$ representation. Instead, it originates from the interplay between the intrinsic zero-energy flat band and the non-Hermitian compression of the surrounding dispersive bands, provided that the effective pseudospin field retains the separable form $\boldsymbol{d}(k)=\left(t_0, 2t_1\cos k, i\gamma\right)$.
	Under this condition, increasing the gain--loss strength drives the dispersive bands toward the flat band without shifting its energy. For an integer pseudospin-$n$ representation, this process culminates in the coalescence of all $2n+1$ bands at zero energy, giving rise to a $(2n+1)$th-order EP. Beyond this critical point, the dispersive bands acquire complex energies while the flat band remains pinned at zero, allowing its compact localized states to persist within the dispersive continuum. The simultaneous occurrence of higher-order exceptional degeneracies and BICs therefore emerges naturally from the same spectral evolution, revealing a direct connection between high-order spectral sensitivity and compact localization in non-Hermitian integer-pseudospin lattices.
	
	\section{Summary}
	
	Our analysis shows that the Bloch Hamiltonian $H(k)=\boldsymbol{d}(k)\cdot\boldsymbol{S}$ provides a general setting for non-Hermitian pseudospin systems, where integer pseudospin representations intrinsically host a zero-energy flat band.
	 Non-Hermiticity compresses the dispersive spectrum toward the intrinsic zero-energy flat band while preserving the compact localization of its associated states, thereby embedding them into the complex continuum beyond the critical point. An additional longitudinal Hermitian contribution
	 preserves compact localization but restricts spectral
	 embedding from a finite interval of non-Hermiticity
	 to isolated parameter values. Its extension to arbitrary integer pseudospin lattices yields a systematic hierarchy of higher-order EPs of order $2n+1$.

	 \section{ Acknowledgments }
	 
	 This work is supported by the National Natural Science Foundation of China (Grant No. 12475021).


\begin{thebibliography}{999}
	\bibitem{LeykamAdvPhysX18}D. Leykam, A. Andreanov, and S. Flach, Artificial flat band systems: from lattice models to experiments, Adv. Phys. X \textbf{3}, 1473052 (2018).
	
	\bibitem{LeykamAPLPhotonics18}D. Leykam and S. Flach, Perspective: Photonic flatbands, APL Photonics \textbf{3}, 070901 (2018).
	
	\bibitem{FlachEurophysLett14}S. Flach, D. Leykam, J. D. Bodyfelt, P. Matthies, and A. S. Desyatnikov, Detangling flat bands into Fano lattices, Europhys. Lett. \textbf{105}, 30001 (2014).
	
	\bibitem{XiaoPhysRevLett17}Y.-X. Xiao, G. Ma, Z.-Q. Zhang, and C. T. Chan, Topological Subspace-Induced Bound State in the Continuum, Phys. Rev. Lett. \textbf{118}, 166803 (2017).
	
	\bibitem{GneitingPhysRevB18}C. Gneiting, Z. Li, and F. Nori, Lifetime of flatband states, Phys. Rev. B \textbf{98}, 134203 (2018).
	
	\bibitem{CaoNature18}Y. Cao, V. Fatemi, S. Fang, K. Watanabe, T. Taniguchi, E. Kaxiras, and P. Jarillo-Herrero, Unconventional superconductivity in magic-angle graphene superlattices, Nature \textbf{556}, 43 (2018).
	
	\bibitem{YankowitzScience19}M. Yankowitz, S. Chen, H. Polshyn, Y. Zhang, K. Watanabe, T. Taniguchi, D. Graf, A. F. Young, and C. R. Dean, Tuning superconductivity in twisted bilayer graphene, Science \textbf{363}, 1059 (2019).
	
	\bibitem{BalentsNatPhys20}L. Balents, C. R. Dean, D. K. Efetov, and A. F. Young, Superconductivity and strong correlations in moiré flat bands, Nat. Phys. \textbf{16}, 725 (2020).
	
	\bibitem{LeykamPhysRevA12}D. Leykam, O. Bahat-Treidel, and A. S. Desyatnikov, Pseudospin and nonlinear conical diffraction in Lieb lattices, Phys. Rev. A \textbf{86}, 031805 (2012).
	
	\bibitem{LeykamPhysRevB13}D. Leykam, S. Flach, O. Bahat-Treidel, and A. S. Desyatnikov, Flat band states: Disorder and nonlinearity, Phys. Rev. B \textbf{88}, 224203 (2013).
	
	\bibitem{CaceresAravenaPhysRevA19}
	G. Cáceres-Aravena and R. A. Vicencio,
	Perfect localization on flat-band binary one-dimensional photonic lattices,
	Phys. Rev. A \textbf{100}, 013803 (2019).
	
	\bibitem{CaceresAravenaPhysRevA20}
	G. Cáceres-Aravena, L. E. F. Foa Torres, and R. A. Vicencio,
	Topological and flat-band states induced by hybridized linear interactions in one-dimensional photonic lattices,
	Phys. Rev. A \textbf{102}, 023505 (2020).
	
	\bibitem{BermudezPhysRevLett09}A. Bermudez, D. Patanè, L. Amico, and M. A. Martin-Delgado, Topology-Induced Anomalous Defect Production by Crossing a Quantum Critical Point, Phys. Rev. Lett. \textbf{102}, 135702 (2009).
	
	\bibitem{SunPhysRevLett11}K. Sun, Z. Gu, H. Katsura, and S. Das Sarma, Nearly Flatbands with Nontrivial Topology, Phys. Rev. Lett. \textbf{106}, 236803 (2011).
	
	\bibitem{TangPhysRevLett11}E. Tang, J.-W. Mei, and X.-G. Wen, High-Temperature Fractional Quantum Hall States, Phys. Rev. Lett. \textbf{106}, 236802 (2011).
	
	\bibitem{NeupertPhysRevLett11}T. Neupert, L. Santos, C. Chamon, and C. Mudry, Fractional Quantum Hall States at Zero Magnetic Field, Phys. Rev. Lett. \textbf{106}, 236804 (2011).
	
	\bibitem{GuzmanSilvaNewJPhys14}D. Guzmán-Silva, C. Mejía-Cortés, M. A. Bandres, M. C. Rechtsman, S. Weimann, S. Nolte, M. Segev, A. Szameit, and R. A. Vicencio, Experimental observation of bulk and edge transport in photonic Lieb lattices, New J. Phys. \textbf{16}, 063061 (2014).
	
	\bibitem{ZhangPhysRevA19}S. M. Zhang and L. Jin, Flat band in two-dimensional non-Hermitian optical lattices, Phys. Rev. A \textbf{100}, 043808 (2019).
	
	\bibitem{MasumotoNewJPhys12}N. Masumoto, N. Y. Kim, T. Byrnes, K. Kusudo, A. Löffler, S. Höfling, A. Forchel, and Y. Yamamoto, Exciton–polariton condensates with flat bands in a two-dimensional kagome lattice, New J. Phys. \textbf{14}, 065002 (2012).
	
	\bibitem{ChernPhysRevA14}G.-W. Chern, C.-C. Chien, and M. Di Ventra, Dynamically generated flat-band phases in optical kagome lattices, Phys. Rev. A \textbf{90}, 013609 (2014).
	
	\bibitem{VidalPhysRevLett98}J. Vidal, R. Mosseri, and B. Douçot, Aharonov-Bohm Cages in Two-Dimensional Structures, Phys. Rev. Lett. \textbf{81}, 5888 (1998).
	
	\bibitem{KolovskyPhysRevB18}A. R. Kolovsky, A. Ramachandran, and S. Flach, Topological flat Wannier-Stark bands, Phys. Rev. B \textbf{97}, 045120 (2018).
	
	\bibitem{ChalkerPhysRevB10}J. T. Chalker, T. S. Pickles, and P. Shukla, Anderson localization in tight-binding models with flat bands, Phys. Rev. B \textbf{82}, 104209 (2010).
	
	\bibitem{ShuklaPhysRevB18}P. Shukla, Disorder perturbed flat bands: Level density and inverse participation ratio, Phys. Rev. B \textbf{98}, 054206 (2018).
	
	\bibitem{LiuChinPhysB14}Z. Liu, F. Liu, and Y.-S. Wu, Exotic electronic states in the world of flat bands: From theory to material, Chin. Phys. B \textbf{23}, 077308 (2014).
	
	\bibitem{MukherjeePhysRevLett15}S. Mukherjee, A. Spracklen, D. Choudhury, N. Goldman, P. Öhberg, E. Andersson, and R. R. Thomson, Observation of a Localized Flat-Band State in a Photonic Lieb Lattice, Phys. Rev. Lett. \textbf{114}, 245504 (2015).
	
	\bibitem{VicencioPhysRevLett15}R. A. Vicencio, C. Cantillano, L. Morales-Inostroza, B. Real, C. Mejía-Cortés, S. Weimann, A. Szameit, and M. I. Molina, Observation of Localized States in Lieb Photonic Lattices, Phys. Rev. Lett. \textbf{114}, 245503 (2015).
	
	\bibitem{RechtsmanNatPhoton13}M. C. Rechtsman, J. M. Zeuner, A. Tünnermann, S. Nolte, M. Segev, and A. Szameit, Strain-induced pseudomagnetic field and photonic Landau levels in dielectric structures, Nat. Photonics \textbf{7}, 153 (2013).
	
	\bibitem{BiondiPhysRevLett15}M. Biondi, E. P. L. van Nieuwenburg, G. Blatter, S. D. Huber, and S. Schmidt, Incompressible Polaritons in a Flat Band, Phys. Rev. Lett. \textbf{115}, 143601 (2015).
	
	\bibitem{VicencioPobleteAdvPhysX21}
	R. A. Vicencio Poblete,
	Photonic flat band dynamics,
	Adv. Phys. X \textbf{6}, 1878057 (2021).
	
	\bibitem{TaieSciAdv15}S. Taie, H. Ozawa, T. Ichinose, T. Nishio, S. Nakajima, and Y. Takahashi, Coherent driving and freezing of bosonic matter wave in an optical Lieb lattice, Sci. Adv. \textbf{1}, e1500854 (2015).
	
	\bibitem{LeungPhysRevLett20}T.-H. Leung, M. N. Schwarz, S.-W. Chang, C. D. Brown, G. Unnikrishnan, and D. Stamper-Kurn, Interaction-Enhanced Group Velocity of Bosons in the Flat Band of an Optical Kagome Lattice, Phys. Rev. Lett. \textbf{125}, 133001 (2020).
	
	\bibitem{ChaseMayoralPhysRevB23}C. Chase-Mayoral, L. Q. English, N. Lape, Y. Kim, S. Lee, A. Andreanov, S. Flach, and P. G. Kevrekidis, Compact localized states in electric circuit flat-band lattices, Phys. Rev. B \textbf{109}, 075430 (2024).
	
	\bibitem{vonNeumannPhysZ29}
	J. von Neumann and E. P. Wigner,
	Über merkwürdige diskrete Eigenwerte, Phys. Z. \textbf{30}, 465 (1929).
	
	\bibitem{DoelemanNatPhotonics18}H. M. Doeleman, F. Monticone, W. den Hollander, A. Alù, and A. F. Koenderink, Experimental observation of a polarization vortex at an optical bound state in the continuum, Nat. Photonics \textbf{12}, 397 (2018).
	
	\bibitem{YuOptica19}Z. Yu, X. Xi, J. Ma, H. K. Tsang, C.-L. Zou, and X. Sun, Photonic integrated circuits with bound states in the continuum, Optica \textbf{6}, 1342 (2019).
	
	\bibitem{KoshelevScience20}K. Koshelev, S. Kruk, E. Melik-Gaykazyan, J.-H. Choi, A. Bogdanov, H.-G. Park, and Y. Kivshar, Subwavelength dielectric resonators for nonlinear nanophotonics, Science \textbf{367}, 288 (2020).
	
	\bibitem{KangNatRevPhys23}M. Kang, T. Liu, C. T. Chan, and M. Xiao, Applications of bound states in the continuum in photonics, Nat. Rev. Phys. \textbf{5}, 659 (2023).
	
	\bibitem{FengPhysRevAppl23}Z. Feng, Y. Liu, X. Xi, L. Wang, and X. Sun, Gigahertz Phononic Integrated Circuits Based on Overlay Slot Waveguides, Phys. Rev. Applied \textbf{19}, 064076 (2023).
	
	\bibitem{HuangAdvSci22}L. Huang, B. Jia, Y. K. Chiang, S. Huang, C. Shen, F. Deng, T. Yang, D. A. Powell, Y. Li, and A. E. Miroshnichenko, Topological Supercavity Resonances in the Finite System, Adv. Sci. \textbf{9}, 2200257 (2022).
	
	\bibitem{HuangNatCommun21} L. Huang, Y. K. Chiang, S. Huang, C. Shen, F. Deng, Y. Cheng, B. Jia, Y. Li, D. A. Powell, and A. E. Miroshnichenko, Sound trapping in an open resonator, Nat. Commun. \textbf{12}, 4819 (2021).
	
	\bibitem{CapassoNature92}F. Capasso, C. Sirtori, J. Faist, D. L. Sivco, S.-N. G. Chu, and A. Y. Cho, Observation of an electronic bound state above a potential well, Nature \textbf{358}, 565 (1992).
	
	\bibitem{ChenNewJPhys16}Y. Chen, Z. Shen, X. Xiong, C.-H. Dong, C.-L. Zou, and G.-C. Guo, Mechanical bound state in the continuum for optomechanical microresonators, New J. Phys. \textbf{18}, 063031 (2016).
	
	\bibitem{YangPhysRevLett20}Y. Yang, Y.-P. Wang, J. W. Rao, Y. S. Gui, B. M. Yao, W. Lu, and C.-M. Hu, Unconventional Singularity in Anti-Parity-Time Symmetric Cavity Magnonics, Phys. Rev. Lett. \textbf{125}, 147202 (2020).
	
	\bibitem{RamezaniPhysRevA17}H. Ramezani, Non-Hermiticity-induced flat band, Phys. Rev. A \textbf{96}, 011802 (2017).
	
	\bibitem{MonticonePhysRevLett14}F. Monticone and A. Alù, Embedded Photonic Eigenvalues in 3D Nanostructures, Phys. Rev. Lett. \textbf{112}, 213903 (2014).
	
	\bibitem{MurPetitPhysRevB14}J. Mur-Petit and R. A. Molina, Chiral bound states in the continuum, Phys. Rev. B \textbf{90}, 035434 (2014).
	
	\bibitem{MolinaPhysRevLett12}M. I. Molina, A. E. Miroshnichenko, and Y. S. Kivshar, Surface Bound States in the Continuum, Phys. Rev. Lett. \textbf{108}, 070401 (2012).
	
	\bibitem{CorrielliPhysRevLett13}G. Corrielli, G. Della Valle, A. Crespi, R. Osellame, and S. Longhi, Observation of Surface States with Algebraic Localization, Phys. Rev. Lett. \textbf{111}, 220403 (2013).
	
	\bibitem{FengLightSciAppl26}Z. Feng, L. Jin, and X. Sun, Anti-PT symmetry with bound states in the continuum, Light Sci. Appl. \textbf{15}, 306 (2026).
	
	\bibitem{YinNature20}X. Yin, J. Jin, M. Soljačić, C. Peng, and B. Zhen, Observation of topologically enabled unidirectional guided resonances, Nature \textbf{580}, 467 (2020).
	
	\bibitem{HwangNatCommun21}M.-S. Hwang, H.-C. Lee, K.-H. Kim, K.-Y. Jeong, S.-H. Kwon, K. Koshelev, Y. Kivshar, and H.-G. Park, Ultralow-threshold laser using super-bound states in the continuum, Nat. Commun. \textbf{12}, 4135 (2021).
	
	\bibitem{WeimannPhysRevLett13}S. Weimann, Y. Xu, R. Keil, A. E. Miroshnichenko, A. Tünnermann, S. Nolte, A. A. Sukhorukov, A. Szameit, and Y. S. Kivshar, Compact Surface Fano States Embedded in the Continuum of Waveguide Arrays, Phys. Rev. Lett. \textbf{111}, 240403 (2013).
	
	\bibitem{KoshelevPhysRevLett18}K. Koshelev, S. Lepeshov, M. Liu, A. Bogdanov, and Y. Kivshar, Asymmetric Metasurfaces with High-Q Resonances Governed by Bound States in the Continuum, Phys. Rev. Lett. \textbf{121}, 193903 (2018).
	
	\bibitem{LonghiPhysRevA08}S. Longhi, Transfer of light waves in optical waveguides via a continuum, Phys. Rev. A \textbf{78}, 013815 (2008).
	
	\bibitem{LonghiPhysRevA09}S. Longhi, Optical analog of population trapping in the continuum: Classical and quantum interference effects, Phys. Rev. A \textbf{79}, 023811 (2009).
	
	\bibitem{PlotnikPhysRevLett11}Y. Plotnik, O. Peleg, F. Dreisow, M. Heinrich, S. Nolte, A. Szameit, and M. Segev, Experimental Observation of Optical Bound States in the Continuum, Phys. Rev. Lett. \textbf{107}, 183901 (2011).
	
	
	
    \bibitem{Leykam2017}D. Leykam, S. Flach, and Y. D. Chong, “Flat bands in lattices with non-Hermitian coupling,” Phys. Rev. B 96, 064305 (2017). 
    
     \bibitem{Jin2019}L. Jin, “Flat band induced by the interplay of synthetic magnetic flux and non-Hermiticity,” Phys. Rev. A 99, 033810 (2019). 
     
    
    
    
		
	\bibitem{BerciouxPhysRevA09}D. Bercioux, D. F. Urban, H. Grabert, and W. Häusler, Massless Dirac-Weyl fermions in a $\mathcal{T}_3$ optical lattice, Phys. Rev. A \textbf{80}, 063603 (2009).
	
	\bibitem{DoraPhysRevB11}B. Dóra, J. Kailasvuori, and R. Moessner, Lattice generalization of the Dirac equation to general spin and the role of the flat band, Phys. Rev. B \textbf{84}, 195422 (2011).
	
	\bibitem{BenderPhysRevLett98}
	C. M. Bender and S. Boettcher, Real Spectra in Non-Hermitian Hamiltonians Having PT Symmetry, Phys. Rev. Lett. \textbf{80}, 5243 (1998).
	
	\bibitem{DembowskiPhysRevLett01}
	C. Dembowski, H.-D. Gräf, H. L. Harney, A. Heine, W. D. Heiss, H. Rehfeld, and A. Richter, Experimental Observation of the Topological Structure of Exceptional Points, Phys. Rev. Lett. \textbf{86}, 787 (2001).
	
	\bibitem{HsuNatRevMater16}
	C. W. Hsu, B. Zhen, A. D. Stone, J. D. Joannopoulos, and M. Soljačić, Bound states in the continuum, Nat. Rev. Mater. \textbf{1}, 16048 (2016).
	
	
	
	
\end{thebibliography}
\end{document}